\documentclass[letterpaper, 10 pt, conference]{ieeeconf}  

\IEEEoverridecommandlockouts                              

\title{\LARGE \bf
Deep Reinforcement Learning based Control Design for Aircraft
Recovery from Loss-of-Control Scenario
}

\author{
\authorblockN{Imran Sayyed$^{1}$ \thanks{Imran Sayyed is an MS Scholar at the Department of Aerospace Engineering, Indian Institute of Technology Madras, Chennai, India. Email: \texttt{imransayyed23@gmail.com}}}
\authorblockA{
Department of Aerospace Engineering\\
Indian Institute of Technology Madras\\
Chennai, India\\
imransayyed23@gmail.com}
\and
\authorblockN{Aayush Konar$^{2}$ \thanks{Aayush Konar is an undergraduate student at the Department of Electrical Engineering, Jadavpur University, Kolkata, India. Email: \texttt{aayushk.electrical.ug@jadavpuruniversity.in}}}
\authorblockA{
Department of Electrical Engineering\\
Jadavpur University\\
Kolkata, India\\
aayushk.electrical.ug@jadavpuruniversity.in}
\and
\authorblockN{Nandan Kumar Sinha$^{3}$ \thanks{Dr. Nandan Kumar Sinha is a Professor in the Flight Dynamics Lab at the Department of Aerospace Engineering, Indian Institute of Technology Madras, Chennai, India. Email: \texttt{nandan@ae.iitm.ac.in}}}
\authorblockA{
Department of Aerospace Engineering\\
Indian Institute of Technology Madras\\
Chennai, India\\
nandan@ae.iitm.ac.in}
}

\usepackage{amsmath} 
\usepackage{amssymb} 
\usepackage{graphicx}
\usepackage{float} 
\begin{document}
\maketitle
\thispagestyle{empty}
\pagestyle{empty}
\begin{abstract}

 Loss-of-control (LOC) remains a leading cause of fixed-wing aircraft accidents, especially in post-stall and flat-spin regimes where conventional gain-scheduled or logic-based recovery laws may fail. This study formulates spin-recovery as a continuous-state, continuous-action Markov Decision Process and trains a Proximal Policy Optimization (PPO) agent on a high-fidelity six-degree-of-freedom F-18/HARV model that includes nonlinear aerodynamics, actuator saturation and rate coupling. A two-phase potential-based reward structure first penalizes large angular rates and then enforces trimmed flight. After 6,000 simulated episodes, the policy generalities to unseen upset initializations. Results show that the learned policy successfully arrests the angular rates and stabilizes the angle of attack. The controller performance is observed to be satisfactory for recovery from spin condition which was compared with a state-of-the-art sliding mode controller. The findings demonstrate that deep reinforcement learning can deliver interpretable, dynamically feasible manoeuvres for real-time loss of control mitigation and provide a pathway for flight-critical RL deployment.

\end{abstract}


\section{INTRODUCTION}

When considering the reliability and safety of aircraft in uncertain circumstances, Loss of Control (LOC) still remains one of the major concerns. Aircraft flight is accompanied by complex and highly nonlinear phenomena that involve underactuated behavior, dynamic couplings between multiple states and nonlinear aerodynamics. In addition to this, the flight envelope constraints, modeling inaccuracies during flight and environmental uncertainties make it even more difficult to have a controlled flight\cite{r1}. Aircraft spin is one such scenario where an aircraft loses altitude rapidly while being fully stalled. An aircraft can become stuck in a near-horizontal condition, a phenomenon known as a flat spin. In this scenario, one wing of the aircraft stalls more than the other, causing it to spin like a disc and lose altitude rapidly. The elevator authority is minimized in the scenario since the nose remains up in a near horizontal position\cite{r2}. From a nonlinear dynamics perspective, a spin condition can be interpreted as an attractive limit cycle in the aircraft's state space where the system undergoes and stays in an uncontrollable oscillatory condition due to strong dynamic coupling between inertial and aerodynamic forces.

It has been discussed in the literature that military aircraft have more spin susceptibility\cite{stough1985stall} than conventional transport aircraft. There are a variety of spin recovery techniques, most common being the PARE (Power idle, Aileron neutral, Rudder opposite and Elevator forward) technique, which is a manual piloted technique. Automatic spin recovery techniques involve pre-programmed controllers that can access the aircraft states in real time and generate required control inputs according to the demand of spin recovery logic embedded in the controller. 

The motion of a rigid aircraft is characterized in six degrees of freedom, i.e., three translation motions and three rotation motions. The dynamics are strongly nonlinear and coupled, and hence it is a challenging task to develop a controller for the aircraft\cite{r7}. Real systems typically also have input saturation limits. These actuator limits lead to control saturation, and the controller does not provide the desired output. This input saturation with uncertainties and disturbances combined can frequently result in loss of control\cite{r3}. In this paper, a Deep Reinforcement Learning (DRL) based control strategy is discussed for aircraft spin recovery. In contrast to classical fixed-logic controllers, an RL agent learns to convert multi-dimensional aircraft state data to the optimal control action through trial-and-error during simulation\cite{doi:10.2514/1.G001739}. The RL policy learns in a high-fidelity flight dynamics environment featuring nonlinear aerodynamic effects, control constraints, and flat spin dynamics; all these states and parameters can be made available to RL agent for accessing through observations. Through repeated interaction with this environment, the agent acquires multiple recovery manoeuvres; these recovery manoeuvres can be similar to traditional automatic controllers or may even be non-intuitive, which are significantly different from logic generated by the automatic controllers\cite{Sutton2018}.

In this work, a spin recovery problem is solved using Reinforcement Learning algorithm known as Proximal Policy Optimization (PPO).  Section II introduced a nonlinear six-degree-of-freedom simulation model of a high-angle-of-attack research vehicle along with actuator saturation constraints. Section III described the key concepts of Reinforcement Learning and provided an introduction to PPO. Section IV describes the implementation of PPO in the Custom F18/HARV RL environment, the design of the reward for the spin-recovery problem, and the training and testing of the agent. Section V presents the simulation results and detailed discussions about the PPO-based recovery policy and performance achieved by the RL controller. Section VI concludes the study, summarizing the key findings and outlining future scope and extensions such as incorporating recurrent networks, safety-constrained learning, and broader envelope-protection objectives for next-generation flight-control systems.

\section{F18/HARV FLight Model Description}
This work considers the nonlinear flight dynamics model of F18/HARV to investigate the nonlinear aerodynamic phenomena. The 6DOF rigid body flight system mathematical model is described below:

\begin{small}
\begin{align}
\dot{V} &= \frac{1}{m} \left[ T_m \eta \cos \alpha \cos \beta - \frac{1}{2} \rho V^2 S C_D(\alpha, q, \delta_e) - mg \sin \gamma \right] \\
\dot{\alpha} &= q - \frac{1}{\cos \beta} \left[ (p \cos \alpha + r \sin \alpha) \sin \beta \right] \notag \\
& \quad + \frac{1}{m V} \left\{ T_m \eta \sin \alpha + \frac{1}{2} \rho V^2 S C_L(\alpha, q, \delta_e) - mg \cos \mu \cos \gamma \right\} \\
\dot{\beta} &= \frac{1}{m V} \left[ - T_m \eta \cos \alpha \sin \beta + \frac{1}{2} \rho V^2 S C_Y(\alpha, \beta, p, r, \delta_e, \delta_a, \delta_r) \right] \notag \\
& \quad + mg \sin \mu \cos \gamma + (p \sin \alpha - r \cos \alpha) \\
\dot{p} &= \frac{I_y - I_z}{I_x} qr + \frac{1}{2 I_x} \rho V^2 S b C_l(\alpha, \beta, p, r, \delta_e, \delta_a, \delta_r) \\
\dot{q} &= \frac{I_z - I_x}{I_y} pr + \frac{1}{2 I_y} \rho V^2 S c C_m(\alpha, q, \delta_e) \\
\dot{r} &= \frac{I_x - I_y}{I_z} pq + \frac{1}{2 I_z} \rho V^2 S b C_n(\alpha, \beta, p, r, \delta_e, \delta_a, \delta_r) \\
\dot{\phi} &= p + q \sin \phi \tan \theta + r \cos \phi \tan \theta \\
\dot{\theta} &= q \cos \phi - r \sin \phi \\
\dot{\psi} &= \frac{q \sin \phi + r \cos \phi}{\cos \theta} \\
 \dot{Z} &= - V \sin \gamma
\end{align}
\end{small}

\begin{small}
\begin{align}
\dot{\mu} &= \sec \beta (p \cos \alpha + r \sin \alpha) 
+ \frac{1}{m V} \bigg\{ 
\frac{1}{2} \rho V^2 S C_L \tan \beta \\ \notag
& \quad + T_m \eta \sin \alpha 
+ \sin \mu \tan \gamma 
- mg \cos \mu \cos \gamma \tan \beta \\ \notag
& \quad + \frac{1}{2} \rho V^2 S C_Y \cos \mu \tan \gamma 
\bigg\}
\end{align}

\begin{align}
\dot{\gamma} &= \frac{1}{m V} \left\{ T_m \eta \left( \sin \alpha \cos \mu + \cos \alpha \sin \beta \sin \mu \right) \right. \notag \\
& \quad \left. - \frac{1}{2} \rho V^2 S C_L \cos \mu - mg \cos \gamma - \frac{1}{2} \rho V^2 S C_Y \sin \mu \right\} 
\end{align}

\begin{align}
\dot{\chi} &= \frac{1}{m V \cos \gamma} \left\{ T_m \eta \left( \sin \alpha \sin \mu - \cos \alpha \sin \beta \cos \mu \right) \right. \notag \\
& \quad \left. + \frac{1}{2} \rho V^2 S \left( C_L \sin \mu + C_Y \cos \mu \right) \right\} 
\end{align}
\end{small}

Although multiple variables in this model are used to investigate the aircraft's performance, only eight variables are independent, which affect the dynamics of the aircraft. It can also be noted that only four actuators are available in the model, yet they must regulate all eight variables; the aircraft model, therefore, is underactuated.

The model formed by all the equations, (1) to (13), can be simply represented as a nonlinear state-space model in (14):

\begin{align}
    \dot{x} = f(x,t) + g(x,t) sat(u)
\end{align}
  
Where 

$$
x(t) = [V, \alpha, \beta, p, q, r, \mu, \gamma] \
$$

$$
u(t) = [\eta, \delta _e, \delta _a, \delta _r ]
$$

Because rank($g(x)) = 4 < 8$ making the system underactuated. 

Here, the $sat(u)$ function models the constraints on the input term. 

\[
sat(u) =
\left\{
\begin{array}{ll}
  u_{max} & \text{if } u > u_{max} \\
  u & \text{if } u_{min} \leq u \leq u_{max} \\
  u_{min} & \text{if } u < u_{min}
\end{array}
\right.
\]

Here, the throttle $\eta$ can only be varied manually and therefore will not be directly considered in the input vector of the system. It is kept at a constant value in this work. Thus, it becomes $u \equiv [\delta _e, \delta _r, \delta _a],$  $u \in \mathbf{R}^3$.

\section{Reinforcement Learning}

Reinforcement Learning (RL) is a branch of Machine Learning that calculates and optimises the actions (control inputs in our case) based on Bellman Optimality Conditions given that the process is a Markov Decision Process (MDP). It uses a mechanism and trial and error and reward feedback to calculate the optimal actions such that the cumulative reward can be maximized\cite{Sutton2018}. This is usually termed as exploration and exploitation. It mainly includes an agent and an environment; the agent interacts with the environment through different actions and gets a reward in return.

The goal of the agent is to learn a policy such that the cumulative reward obtained is maximized. This reward feedback mechanism distinguishes RL from other types of machine learning processes, which have supervised and unsupervised Learning. It tries to learn a policy i.e. $\pi(a|s)$ and optimizes the actions such that the policy becomes $\pi ^* (a|s)$, which is known as the optimal policy\cite{Lillicrap2016DDPG}. The policy here can be interpreted as a plan or a control law that gives the probabilities of all actions for a particular state, and the action is taken using those probabilities. 

Thus, finally agent crawls between states, which are guided through a value function maximized by actions given by the optimal policy. This can be achieved in multiple ways and therefore there exist multiple methods, such as value-based methods, policy-based methods, also some other types of methods which may or may not use these methods discussed earlier. These methods have been successfully applied in multiple domains, which include robotics, games, networked systems and autonomous systems\cite{Garcia2015SafeRLSurvey}.

\subsection{Proximal Policy Optimization (PPO)}

Proximal Policy Optimization (PPO) is an on-policy reinforcement learning algorithm, which means the same policy is employed for decision-making during training as well as data collection (i.e., the behaviour policy and the policy under use are the same). Due to this reason, PPO does not use a replay buffer like off-policy algorithms like DQN and DDPG\cite{Schulman2017PPO}.

PPO utilizes a stochastic policy and updates it with data sampled via on-policy rollouts. At every training iteration, the agent samples a batch of transitions with the current policy and subsequently updates the policy using this batch.

\begin{figure}[H]
    \centering
    \includegraphics[width=1\linewidth]{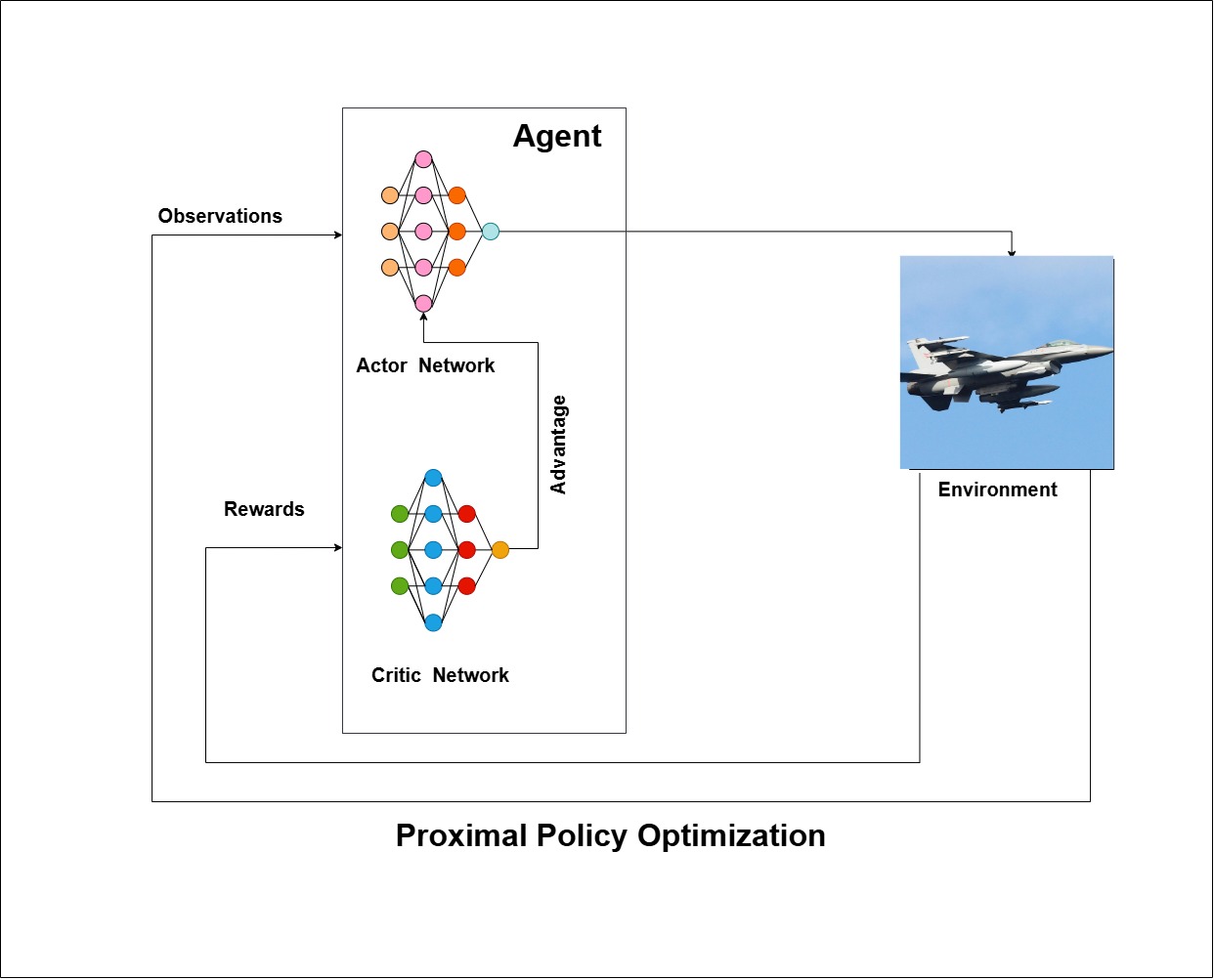}
    \caption{Proximal Policy Optimization Structure for Flight Control}
    \label{fig:enter-label}
\end{figure}

An entropy bonus is added to the loss function to promote improved exploration. PPO alleviates the computational issues of Trust Region Policy Optimization (TRPO) by having a more realistic surrogate objective function for the actor loss\cite{Schulman2015TRPO}. Two formulations of this objective exist: one with a KL-divergence penalty, and one using a clipped probability ratio, the latter being the most popular. The clipped surrogate objective prevents too-large policy updates and allows stable training. It uses an actor-critic network architecture just like DDPG as shown in Figure 1.
Proximal Policy Optimization (PPO) is an RL technique introduced by \cite{Schulman2017PPO}, which is a state-of-the-art algorithm that gives very stable and efficient policy updates. Traditional policy gradient methods often do large updates in policies and thus often make the learning unstable.

The most important components of an RL MDP framework is described by the tuple $(s, a, p, r, \gamma)$ where $s$ represents the state of the system, $a$ represents action, $p$ represents the probability of transition of one state to another, $r$ represents the reward and $\gamma$ represents that discount factor which helps us tune short term and long term rewards.

PPO uses an objective function that penalizes the updates of the policy if it updates the policy beyond a threshold value. The objective function is defined by (16) :
\begin{equation}
r_t(\theta) = \frac{\pi_\theta(a_t \mid s_t)}{\pi_{\theta_{\text{old}}}(a_t \mid s_t)}
\end{equation}

\begin{equation}
    L^{clipped} = \mathbb{E} [ min (r_t(\theta). A_t, clip(r_t(\theta), 1- \epsilon, 1 + \epsilon).a_t )] 
\end{equation}

Here $E$ is Expectation, $\epsilon$ is the clipping range, $A_t$ represents the advantage value\cite{Schulman2017PPO}.

The algorithm of PPO can be described as Algorithm 1 in this work:
\begin{center}
\small
\centering
\begin{tabular}{@{}l@{~}p{7 cm}@{}}
\hline
\textbf{Algorithm 1} & \textbf{PPO algorithm} \\
\hline
1:  & Initialize the actor network $\pi_\theta$ and the value network $V_\phi$ \\
2:  & \textbf{for} Iteration $= 1, 2, \dots, M$ \textbf{do} \\
3:  & \hspace{1em} \textbf{for} Episode $= 1, 2, \dots, N$ \textbf{do} \\
4:  & \hspace{2em} \textbf{for} $t = 1, 2, \dots, T$ \textbf{do} \\
5:  & \hspace{3em} Using policy $\pi_\theta$ choose an action $a_t$ \\
6:  & \hspace{3em} If termination conditions hold true, end the episode \\
7:  & \hspace{2em} \textbf{end for} \\
8:  & \hspace{1em} Find advantage values ${A}_1, {A}_2, \dots, {A}_T$ using $V_\phi$ \\
9:  & \hspace{1em} \textbf{end for} \\
10: & \hspace{1em} \textbf{for} Epochs $= 1, 2, \dots, K$ \textbf{do} \\
11: & \hspace{2em} Update actor netwrok using value $L^{\text{CLIP}}$ \\
12: & \hspace{2em} Update value using critic loss\\
13: & \hspace{1em} \textbf{end for} \\
14: & \textbf{end for} \\
\hline
\end{tabular}
\end{center}

\section{Training and Simulation}

A custom RL-based environment for the F18/HARV aircraft model is created using OpenAI Gym\cite{Brockman2016Gym}. The Observation space and action space of the gym environment are defined in Table 1.

\begin{table}[ht]
\footnotesize
\centering
\caption{Observation and Action Space (All Angles in Radians)}
\begin{tabular}{|c|p{2.5cm}|c|c|c|}
\hline
\textbf{Type} & \textbf{Variable (Unit)} & \textbf{Dim.} & \textbf{Range} & \textbf{Ref.} \\
\hline
Obs. 
  & Airspeed $V$ (ft/s) & 1 & $[0, 2000]$ & — \\
  & AoA $\alpha$ (rad) & 1 & $[-0.244, 1.571]$ & $\alpha_{ref}$ \\
  & Sideslip $\beta$ (rad) & 1 & $[-\pi, \pi]$ & $0$ \\
  & Bank angle $\mu$ (rad) & 1 & $[-\pi, \pi]$ & $0$ \\
  & Angular rates $p, q, r$ (rad/s) & 3 & $[-10\pi, 10\pi]$ & $0$ \\
  & Flight path angle $\gamma$ (rad) & 1 & $[-1.745, 1.745]$ & — \\
\hline
Act. 
  & Elevator $\delta_e$ (rad) & 1 & $[-0.436, 0.175]$ & — \\
  & Aileron $\delta_a$ (rad) & 1 & $[-0.436, 0.436]$ & — \\
  & Rudder $\delta_r$ (rad) & 1 & $[-0.524, 0.524]$ & — \\
\hline
\end{tabular}
\normalsize
\end{table}

\subsection{Control of Angles and Angular rates}

To recover the aircraft from a flat spin condition, the angular rates ($p$, $q$ and $r$) of an aircraft should be brought down to lower values; after this to maintain a leveled flight condition, the control of states $\alpha$, $\beta$, and $\mu$ is required. Each variable is responsible for stability in pitch, yaw and roll axes respectively\cite{r1}. From (1) and (12), it can seen that as long as $\alpha$, $\beta$, $\mu$ and angular rates are stabilized, $V$ and $\gamma$ will also be stabilized to finite value, subjected to condition that $\eta$ does not change.

The reward function should be designed to first emphasize angular rate attenuation and, once rates fall below a threshold, prioritize maintaining a leveled flight attitude at the target angles\cite{doi:10.2514/1.G001739}. Designing a continuous reward structure for our RL environment is done in two phases. In the first phase, we only focus on rates and in the second phase, we will make our reward prioritize attitude angles in addition to angular rates. To accelerate the training, the Potential Based Reward Shaping (PBRS) is also used\cite{Ng1999PBRS}.

The reward structure is defined as:

\textit{Phase 1}

When $[p,q,r] > 0.17 \: rad/sec$
\begin{align}
    r_{phase_1} = -|| \omega ||^{2} - w_{p_1} ( |pq| + |qr| + |pr|)
\end{align}

Where:

\begin{align*}
    ||\omega ||  = \sqrt{p^2 + q^2 + r^2} 
\end{align*}

$|pq| + |qr| + |pr|$ term represents the couplings between the three axes, which is also undesirable, so it is penalized as well. ${w_p}_1$ is the weight factor for scaling the weights of the rate penalty and cross-coupling penalty. The transition threshold of $0.17$ rad was empirically chosen based on preliminary simulation trials to distinguish the rate-dominant and attitude-dominant phases of recovery. Below this level, the aircraft angular rates are sufficiently attenuated for the motion to enter a near-linear regime, enabling the controller to prioritize attitude stabilization rather than rate damping.

\textit{Phase 2}

When $[p,q,r] < 0.17 \ rad  /sec$
\begin{align}
    r_{phase_2} &= \ -e_{\alpha}^{2} - w_{p_{21}}e_{\alpha} q  \notag \\
    & \ \ \ \ - w_{p_{22}}(|| \omega ||^{2} +  |pq| + |qr| + |pr|) 
\end{align}

Additional Reward:
           \begin{align*}
               & \text{+5 if } |e_\alpha| < 0.017 \ \ rad \\
               & \text{+3 if } |e_\beta| < 0.017 \ \ rad \\
               & \text{+3 if } |e_\mu| < 0.017 \ \ rad \\
               & - \sum_i action_{i}^2
           \end{align*}

Where:

\begin{align*}
    e_{\alpha} &= \alpha \: - \alpha_{d} \\
    e_{\beta} &= \beta \: - \beta_{d} \\
    e_{\mu} &= \mu \: - \mu_{d}
\end{align*}

$\alpha_d$, $\beta_d$ and $\mu_d$ represents the desired or target values of angle of attack, sideslip angle and bank angle respectively. Continuous rewards are not considered for angle of sideslip $\beta$ and bank angle $\mu$ for simplicity in reward design, as it is expected that they to return to the mean position in the long term due to symmetry and therefore stabilization of $p$ and $r$ through additional rewards should be sufficient.  In phase 2, $w_{p_{22}}$ should be taken small enough so that rate and cross-coupling errors are minimized, but at the same time, it must be ensured that they do not dominate. 

Continuous penalty-based rewards penalize the agent at each step regardless of whether the PPO agent makes good or bad progress. But PBRS-based reward guides the agent to learn a better policy by giving a positive reward feedback when the agent moves towards a target and a negative reward if it moves away from the target\cite{Ng1999PBRS}. PBRS-based rewards do not change the optimal conditions but help in accelerating the training process.

\begin{align*}
    \sigma = -e_{\alpha}^2 - w_{p_{21}}||\omega||^2
\end{align*}

Here $\sigma$ is the reward shaping potential.
From (17) and (18), we get the total reward as:

\begin{align}
    \text{Reward} = r_{phase_1} + r_{phase_2} + \Gamma  \sigma_{t+1} - \sigma_{t}
\end{align}

$\sigma_t$ is the potential of current step, and $\sigma_{t+1}$ is the potential of next step (predicted). $\Gamma$ is the discount factor, a hyperparameter that is taken during PPO training for optimizing the balance between current and future rewards. The same $\gamma$  is used in PBRS reward shaping as well.

\begin{center}
\small
\centering
\begin{tabular}{@{}l@{~}p{7 cm}@{}}
\hline
\textbf{Algorithm 2} & \textbf{F-18 Environment Logic 6 DOF} \\
\hline
1:  & \textbf{Input:} Initial state $x_0$, timestep $\Delta t$, control $a_t$ \\
2:  & Initialize F-18 state vector and aircraft parameters \\
3:  & \textbf{function} \texttt{reset()} \\
4:  & \hspace{1em} Sample target angle of attack $\alpha \in [-5^\circ, 40^\circ]$ \\
5:  & \hspace{1em} Initialize state $x_0 = [V, \alpha, \beta, p, q, r, \dots]$ \\
6:  & \hspace{1em} \textbf{return} initial observation $s_0$ \\
7:  & \textbf{end function} \\
8:  & \textbf{function} \texttt{step}($a_t$) \\
9:  & \hspace{1em} Clip and scale $a_t$ to control limits \\
10: & \hspace{1em} Compute control input $u_t$ using Env actions $a_t$\\
11: & \hspace{1em} Integrate aircraft dynamics: $x_{t+1} = f(x_t,u_t)$ \\
12: & \hspace{1em} Update positions and Euler angles \\
13: & \hspace{1em} Compute reward: Potential-based reward shaping or error-based penalty \\
14: & \hspace{1em} Check termination conditions \\
15: & \hspace{1em} \textbf{return} $(s_{t+1}, r_t, d_t, \text{info})$ \\
16: & \textbf{end function} \\
\hline
\end{tabular}
\end{center}

For simulation, the mathematical model of the flight model of F18/HARV is integrated into a Python OpenAI Gym environment\cite{Brockman2016Gym}.  A custom Gym environment was prepared to model the relevant state observations, action space, and dynamics of the agent. Terminal conditions were set if the aircraft goes out of the flight envelope or shows large deviations from target states. The logic followed by the custom F18/HARV is shown in Algorithm 2. Since the maximum achievable cumulative reward in the environment (from initial simulations) was approximately $+30000$, the terminal penalty for failure was set to $-1000$ to maintain a reasonable contrast between successful and unsuccessful episodes. The bounds of observation space and action space are specified in Table 1.

The observation vector consists of the following states:

\begin{align}
    s = \{ V, \alpha, \beta, p, q, r, \mu, \gamma, \alpha_{d}, \beta_{d}, \mu_{d} \}
\end{align}

Action space is defined as:

\begin{align}
    a = \{\delta_e , \delta_a , \delta_r \}
\end{align}

Simulation is initialised at the spin condition $s_0$:

\begin{align*}
    s_0 = [0.186*1116, 1.2375, 0.0382, -0.6163, 0.1784,\\ -1.4645,  -1.3508, -1.5075, 0.3,0,0]
\end{align*}

The aircraft remains in the spin condition for the initial 30 seconds, after which the RL controller takes charge. This setup was intentionally chosen to clearly demonstrate the uncontrolled spin behavior prior to the initiation of the recovery control.

The weights $w_{p_1}$, $w_{p_{21}}$ and  $w_{p_{22}}$ were kept 0.3, 0.3 and 0.05 respectively. Target value $\alpha_d $ was kept at 0.3 (around 17$^o$).  The control inputs are subject to predefined saturation limits to simulate real-world actuator constraints. The actuator constraints for elevator, aileron and rudder are $(-0.43, 0.17) $, $(-0.43, 0.43)$ and $(-0.52, 0.52)$ radians which is  ($-25^o$,\:$10^o$), ($-25^o$,\:$25^o$) and ($-30^o$,\:$30^o$) respectively. The control objective is to track reference trajectories for key state variables (e.g., $\alpha$, $p$, $q$, $r$ etc.) while ensuring that the system can converge to the desired values \cite{10927436}. 
Training of the RL agent was carried out using Stable-Baselines3(SB3) which is a popular Python-based open-source library for reinforcement learning algorithms\cite{Raffin2021SB3}. Proximal Policy Optimization Algorithm was as it is sample efficient and robust in the training process. The network architecture is an Actor-Critic Network of [256, 128] hidden layers - Tanh configuration each. This network size provided a favorable computational balance, being more efficient than larger architectures while still capturing the nonlinear dynamics and couplings that smaller networks failed to represent in prior simulations. Learning rate and Discount factor were kept at 5e-5 and 0.99 to encourage smooth learning and long-term rewards. The agent was trained for a total of 6000 episodes. Each episode consists of 20000 timesteps, with one timestep being of size 0.01 seconds. Agent training performance was logged to check the reward trend with ongoing training using Tensorboard.
\par
We also compared our controller with state-of-the-art sliding mode controllers.
\begin{figure}[H]
    \centering
    \includegraphics[width=0.9\linewidth]{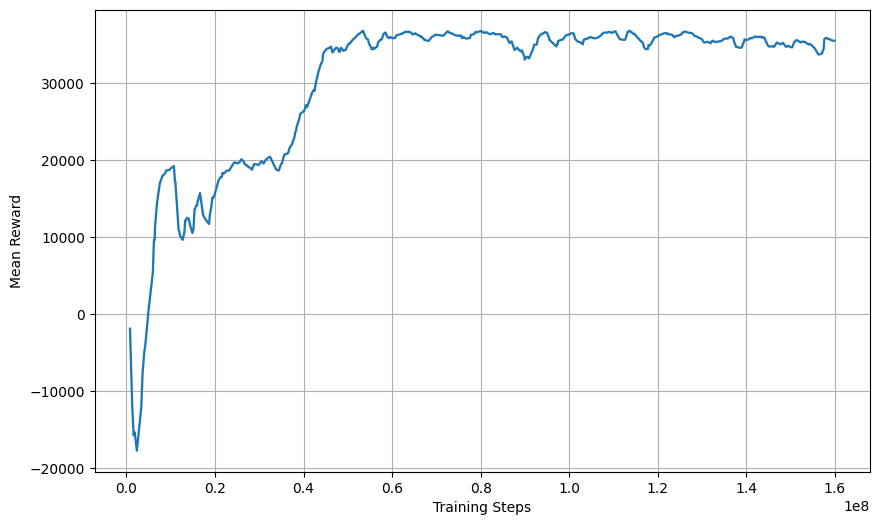}
    \caption{Training progress}
    \label{fig:placeholder}
\end{figure}

\section{Results and Discussions}

After training the model for 6000 episodes, the reward converged as shown in Figure 2, and the results are shown in Figure 3 – Figure 6. It was observed that the RL-based controller was able to arrest angular rates and was also able to stabilize the aircraft to the desired angle of attack position. It can also be observed that the controller managed to arrest the angular rates $p$, $q$ and $r$,  with less overshoots.

\begin{figure}[H]
\centering
\includegraphics[width=1.05\linewidth,trim=25 12 10 20,clip]{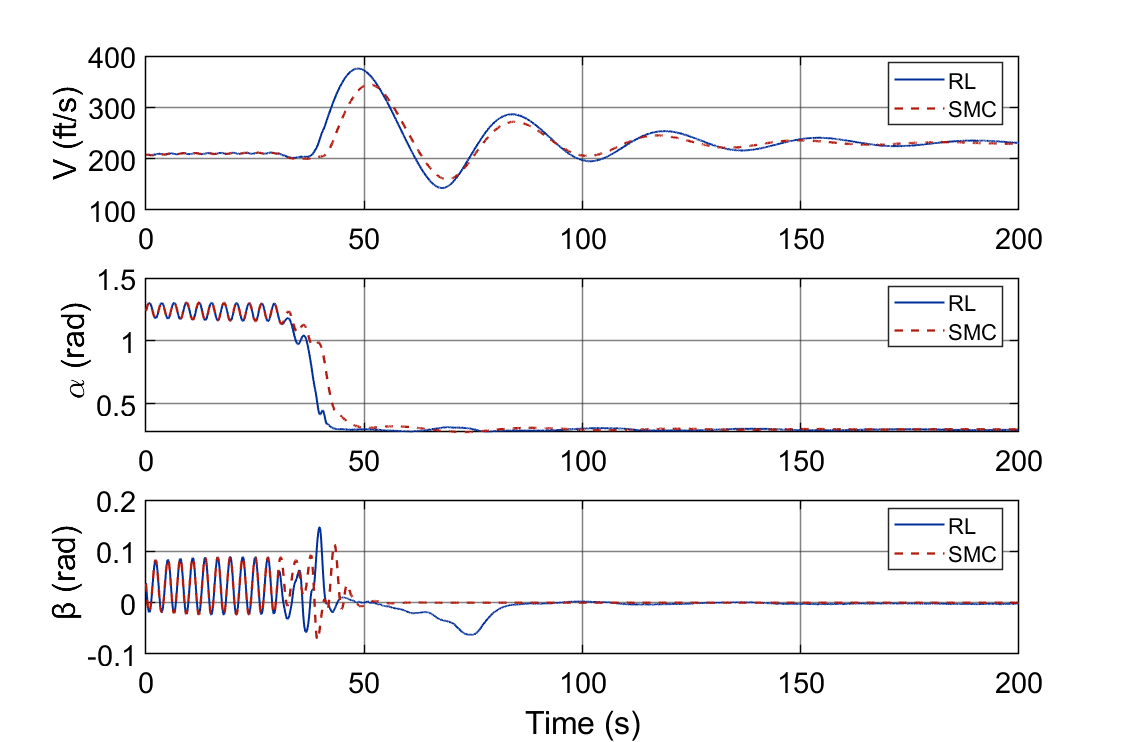}
\caption{(a) Velocity of the Aircraft (ft/sec) (b) Angle of Attack (rads) (c) Angle of Sideslip (rads)}
\label{fig:enter-label}
\end{figure}

\begin{figure}[H]
\centering
\includegraphics[width=1.05\linewidth,trim=25 12 10 20,clip]{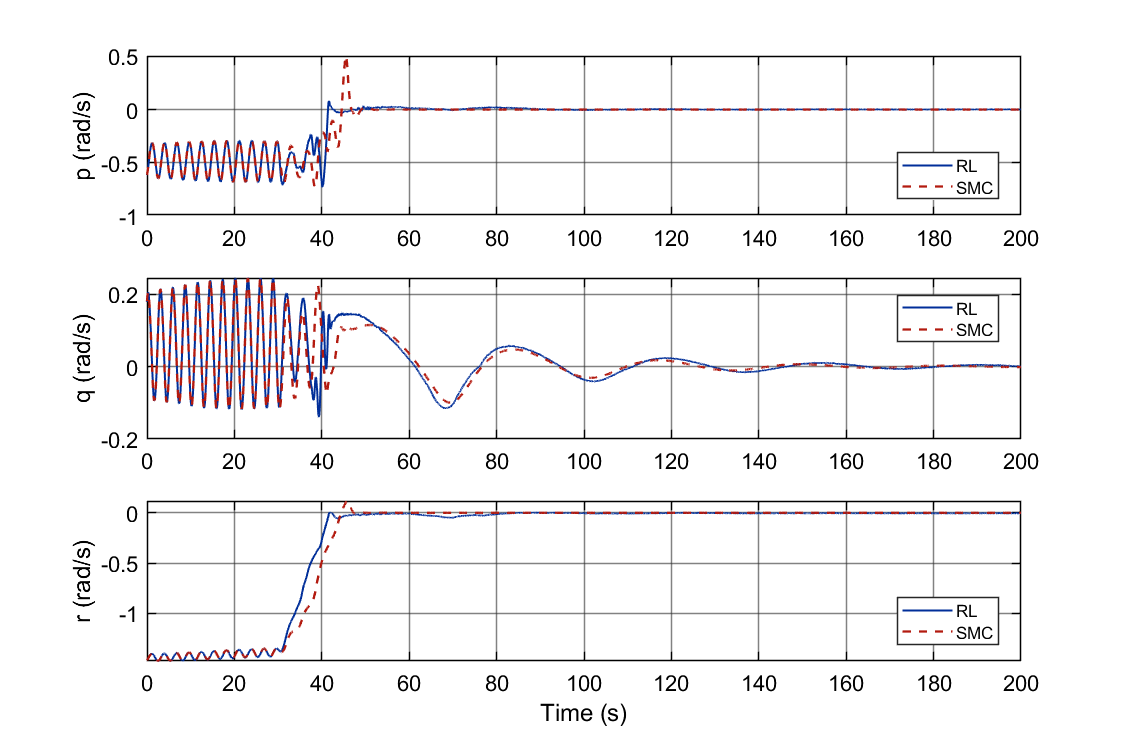}
\caption{(a) Roll rate (rad/sec) (b) Pitch rate (rad/sec) (c) Yaw rate (rad/sec)}
\label{fig:enter-label}
\end{figure}

\begin{figure}[H]
\centering
\includegraphics[width=1.05\linewidth,trim=25 12 10 20,clip]{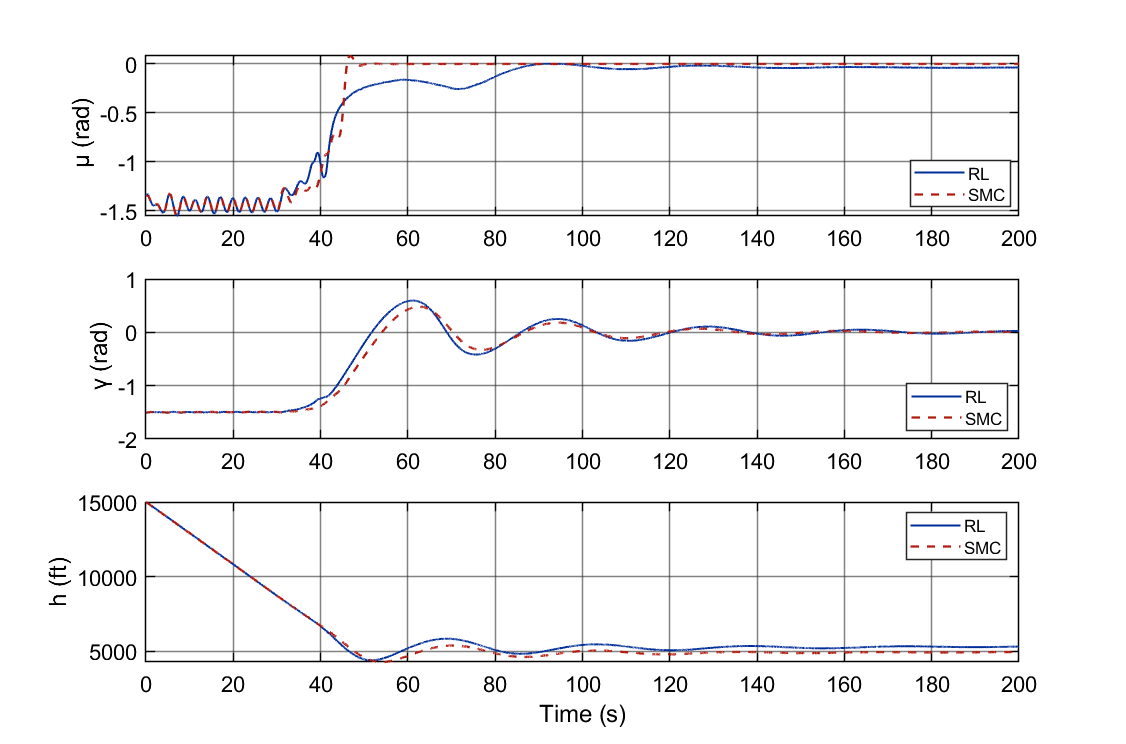}
\caption{(a) Bank angle (rad) (b) Flight path angle (rad) (c) Altitude (ft)}
\label{fig:enter-label}
\end{figure}

\begin{figure}[H]
\centering
\includegraphics[width=1.05\linewidth,trim=25 12 10 20,clip]{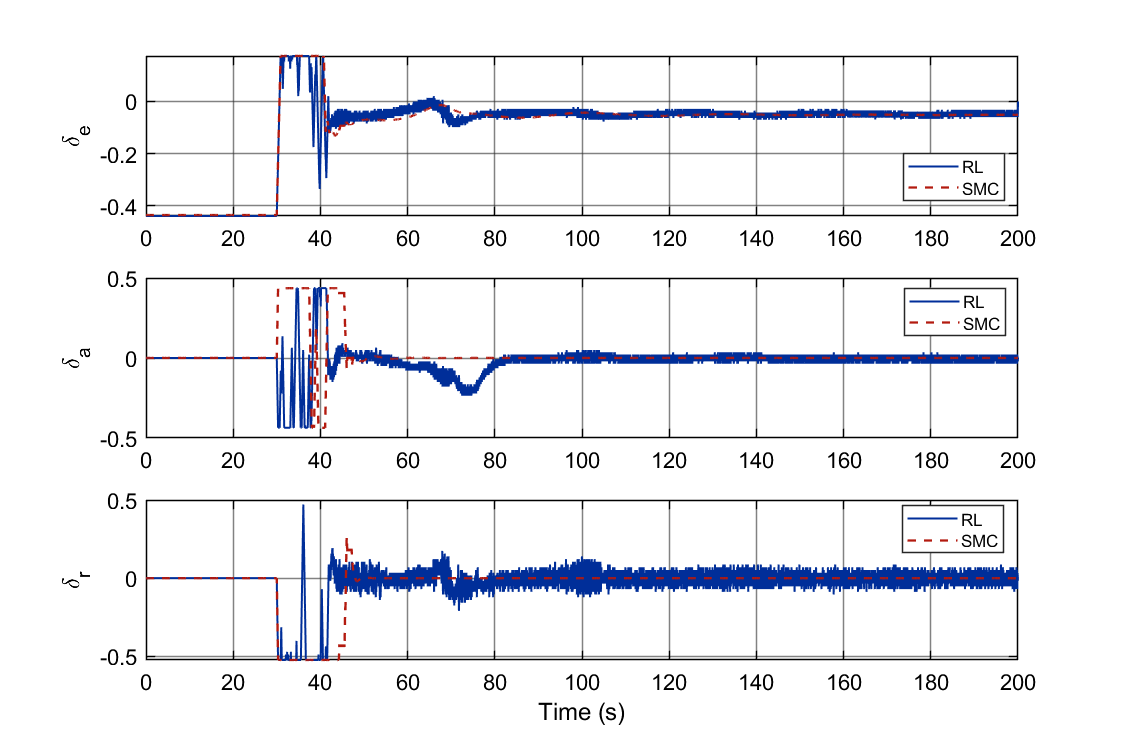}
\caption{(a) Elevator deflection (rad) (b) Aileron deflection (rad) (c) Rudder deflection (rad)}
\label{fig:enter-label}
\end{figure}

 As seen in Figure 3(a) and Figure 3(b) and Figure 5(a), all angle $\alpha$, $\beta$ and $\mu$ are settling desired target values $\alpha_d$, $\beta_d$ and $\mu_d$ respectively. It can be observed from Figure 5(c) that even with the current reward function, which is relatively simple, the trained RL policy shows recovery with the loss of altitude around 4000 ft. (recovery starts around altitude of 8500 ft.).The proposed controller acts faster as it is anticipatory in nature, while other conventional controllers are mostly reactive. This anticipatory action allows improved control performance and quicker recovery, as evidenced from the system response shown in the figures. Chattering seen in Figure 6 is due to not penalizing action rates and can potentially be eliminated by including penalization of control input rate changes in the reward($a_{t+1} - a_t$). Thus, there is scope for improvement in control inputs in reward design by considering their rate of change with respect to time. Overall performance will improve if the reward is designed this way, although it will make the reward structure more complex.

\section{Conclusion}

Uncertainties and perturbations are usually hard to quantify well. This recovery policy in this work was trained with a simple actor-critic architecture. Although this selection maintains the pipeline of training open and reproducible, it hardly touches on what deep-RL toolkits available today can provide for flight-control issues.

Briefly, the straightforward actor-critic network demonstrates that deep-RL can match any highly tuned traditional controller, yet much more sophisticated architectures are still waiting to be unleashed. There is a lot of future scope for RL in nonlinear control, such as recurrent networks such as LSTMs or GRUs could capture the long-term temporal patterns that characterize spins and stalls, attention-based or transformer policies might focus control effort on the most safety-critical states, and safe-RL or Lyapunov-guided methods can embed hard envelope constraints directly into learning. Taking advantage of recurrent memory, attention, explicit models, and safety-conscious objectives is likely to drive recovery times down further, decrease control effort even more, and yield policies that generalize even better to unseen planes and atmospheric states.

\bibliographystyle{unsrt}
\bibliography{ref.bib}


\end{document}